\documentclass{article}
\begin{document}

%\begin{frontmatter}

\title{Cellular Automata for One-Lane Traffic Flow Modelling: Safety
 and Automation}
\author{M.E. L\'{a}rraga \and J.A. del R\'{i}o
%EndAName
Centro de Investigaci\'{o}n en Energ\'{i}a \\
Universidad Nacional Aut\'{o}noma de M\'{e}xico \\
A.P. 34, 62580 Temixco, Mor. \\
M\'{e}xico}

\maketitle

\begin{abstract}
In recent years the modelling of traffic flow using methods from statistical
physics, especially cellular automata models have allowed simulations of
large traffic networks faster than real time. In this paper, we study a
probabilistic cellular automaton model for microsimulations of traffic flow
of automated vehicles in highways. This model describes single-lane traffic
flow on a ring. We study the equilibrium properties by including a parameter
of safe distance in the model and calculate the so-called fundamental
diagrams (flow vs. density graph) considering parallel dynamics. This is
done numerically by computer simulations of the model and by means of an
analysis of speed variance.
\end{abstract}

%\begin{keyword}
Traffic Flow Models Microsimulation  Cellular Automata  
Simulation Modeling
%\end{keyword}

%\end{frontmatter}

\section{Introduction}

In recent years, most industrialized societies have started to see the
limits of the growth of urban traffic. The traffic demand in metropolitan
areas has largely exceeded the vehicular capacity and we face the problem of
increasing pollution and growing frequency of accidents. Typical remedies
such as the expansion of road way systems or road improvement do not work
well anymore. As a consequence, many metropolitan areas seek to solve this
problem by constructing additional roadway, but this is undesirable due to
political and environment objections or impractical because of the high
costs of construction. Thus, the consistent management of large, distributed
man-made transportation systems has become more and more important for
planning and prediction of traffic.

Computer simulations as a means of evaluating, planning, and controlling
traffic systems have gained considerable importance. Lately, more and more
work has been done to consider the dynamical aspects of the traffic system.
Specially, the micro-simulation work has taken great advantage of models
developed in classical research areas such as physics, mathematics, and
computer science. Computer models can be used to simulate the influence of
governmental actions like road-pricing or building of new highways 
\cite{TRA01}, \cite{n2}, \cite{n3}, in solving the traffic problems.
Moreover, output data of traffic models can be used as an input to climate
models for calculating traffic-caused air pollution and the dependency of
this pollution on time-limits.

For this reason, in the last few years there has been considerable interest
in the investigation of traffic flow using methods and models from
statistical physics, especially the cellular automata (CA) models have
allowed the large-scale simulations of large traffic networks faster than
real time \cite{TRA01}. These kind of models are so-called \emph{particle
hopping models}. Recently, particle hopping models using CA has emerged as a
very promising alternative to describe the traffic flow \cite{a1}, \cite{a2}%
, \cite{a3}, \cite{a4}.

Cellular automata are alternatives to Differential Equations in an attempt
to model transportation systems. CA's are dynamical systems in which space
and time are discrete. A cellular automaton consists of a regular grid of
cells, each of which can be in one of a finite number of possible $k$
states, updated synchronously in discrete time steps according to a local,
identical interaction rule. The state of a cell is determined by the
previous states of the surrounding neighborhood of cells. Thus, CA models
have the desirable capacity to capture micro-level dynamics and relate these
to macro-level behavior. Also they are capable of representing individual
vehicle interactions and relating these interactions to macroscopic traffic
flow metrics, such as throughput, time travel, vehicle speed, etc. In
addition, CA models are suitable to represent both single and multi-lane
traffic in an easy form, which is particularly crucial for the modeling of
highways \cite{a2}. The technical difference between car-following and CA
models for traffic flow is that in the latter, space and time are discrete,
whereas in the car-following models, they are continuous. Simulations of
car-following models use discrete time but the space is continuous.

The objective of this paper is to explore a new minimal CA\ model for the
high-speed simulation of microscopic traffic alternatives. We use particle
hopping models as a starting point for this investigation because their
highly discrete nature reduces the number of free parameters. Here, we
present a new extension of a CA model for a single-lane highway on a ring
built on the pioneering work of Nagel (NS) and his colleges \cite{nag}. The
NS model has been applied to the project TRANSIMS on transportation
simulation. This model consider four rules to update the state of the road
involving braking, acceleration, stochastic driver reaction, and the car
movement. The NS model decelerates a vehicle if its distance from the
vehicle in the front is less than its current speed but does not consider
the speed of the vehicle ahead. By including a parameter of safe distance we
determine the distance that one vehicle must maintain with respect to the
vehicle in the front by taking their speeds into account. The knowledge of
this speed is a property of some kind of automated vehicles. We analyze the
relationship between the flow and the density, the so-called fundamental
diagram, and we find that the new model works better under the condition of
high traffic density. By analyzing the variance of speed and its
relationship with the flow, we show that the behavior of this variance can
have important consequences for the design and control of traffic flow. We
think that knowledge of this relationship can be very useful for analyzing 
\emph{automated system} especially for vehicles equipped with infrared
sensors to determine the distance and the velocity from other vehicle.
Recently such vehicles have been used to study an automated traffic \cite
{sci}. In the following we will refer as automated vehicles to this kind of
vehicles.

The paper is organized as follows. In Sect. 2, we give a brief description
of our model which modifies the existing NS model by an inclusion of a
safety criterion. In Section 3, the results of computer simulations from the
new model are presented, by analyzing fundamental diagrams for different
values of the safety parameter. In this section, we also show that a
suitable safety factor produces a lesser speed variance. Finally, Sect. 4
concludes with a short summary and discussion of our findings.

\section{Definition of the model}

Our model is a probabilistic cellular automaton. It consists of $N$ cars
moving on a one-dimensional lattice of $L$ cells with periodic boundary
conditions (the number of vehicles is conserved). Space and time are
discrete and hence also the velocities. Each cell is either empty, or it is
occupied by just one vehicle (see Fig. 1) with a discrete velocity $v,$ with
the same limt speed. In the present paper, velocity ranges are from $%
0,\ldots ,v_{max}=5,$ the limit speed. This limit speed may be different
depending of the kind of vehicle under consideration, trucks, cars, etc.
however, here will use the same maximum velocity for all the vehicles, then
we consider only one type of car. The velocity is equivalent to the number
of sites that a vehicle advances in one time step --- provided that there
are no obstacles ahead. Vehicles only move in one direction. The length of a
cell is around $7.5m$ $(\Delta x)$, which is interpreted as the length of a
vehicle plus the distance between vehicles in a jam, but can be suitably
adjusted with respect to the problem under consideration. One time step $%
t\rightarrow t+1$ ($\Delta t$) lasts 1 s, which is of the order of the
reaction time of humans. Then, $v=1$ means $27$ Km/h in real units.

Let $v_{i}$ and $x_{i}$ denote the current velocity and position
respectively of the vehicle $i,$ and $v_{p}$ and $x_{p}$ be the velocity and
position respectively of the vehicle ahead (preceding vehicle) in a fixed
time. We denote by $d_{i}:=x_{p}-x_{i}-1$, the distance (number of empty
cells) in front of the vehicle in the position $x_{i}$. In the context of CA
models proper units are often omitted due to the given discretization of
space and time. Thus, the proper units for the model would be: $[d]=$number
of cells, $[v]=$number of cells per time step, $[t]=$number of time steps,
etc. For that reason we will use $v<d$ instead of $v<d/\Delta t$, because $%
\Delta t=1$. We have the following set of rules, which are applied in \emph{%
parallel} (simultaneously) for all cars:

S1: Acceleration ($\mathcal{A}$). If $v_{i}<v_{max}$, the velocity of the
car $i$ is increased by one, i.e.,

$\ \ \ \ \ \ \ \ \ \ \ \ v_{i}\rightarrow min(v_{i}+1,v_{max})$

This rule simulates that all the drivers like to reach the maximum velocity\ 

S2: Randomization ($\mathcal{R}$). If $v_{i}>0$, the velocity of the car $i$%
\ is decreased randomly by one unity with probability $R$, i.e.,

$\ \ \ \ \ \ \ \ \ \ \ \ v_{i}\rightarrow max(v_{i}-1,0)$ with probability $%
R.$

\noindent This rule represents all the random situations, such as the random
driver behavior or the random highway physical state.

S3: Deceleration ($\mathcal{D}$). If $(d_{i}+(1-\alpha ))\cdot v_{p}<v_{i}$,
the velocity of the car $i$\ is reduced to ($d_{i}+(1-\alpha )\cdot v_{p})$,
i.e., the new velocity of the vehicle $i$ is:

$\ \ \ \ \ \ \ \ \ \ \ \ v_{i}\rightarrow min(v_{i},(d_{i}+(1-\alpha )\cdot
v_{p}))$\noindent .

\noindent The term $(d_{i}+(1-\alpha )\cdot v_{p})$ represents the driving
schemes respecting the safety distance (gap) and this distance is determined
by the safe distance\ parameter $\alpha .$ The physical meaning of $\alpha $
will be discussed latter.

S4: Vehicle movement ($\mathcal{M}$). Each car is moving forward according
to its new velocity determined in steps 1-3, i.e.,

$\ \ \ \ \ \ \ \ \ \ x_{i}\rightarrow x_{i}+v_{i}$

The ''rules'' $S1,\ldots ,S3$ allow to obtain the new velocities. Then the
vehicles are moved ($S4).$ Thus, one has to divide the update in two parts,
the first part is to obtain the new velocities while the second part is the
movement of cars.

Step 1 reflects the general tendency of the drivers to drive as fast as
possible without crossing the maximum speed limit. The randomization in step
2 \ takes into account the different behavioral patterns of the individual
drivers, specially, nondeterministic acceleration as well as overreaction
while slowing down. This is important for the spontaneous formation of
traffic jam. Step 3 is intended to avoid collision between the cars. We use
a parallel updating scheme since it takes into account the reaction time and
can lead to a chain of overreactions. Suppose, a car slows down due to the
randomization step, and if the density is large enough then this might force
the following cars to brake in the deceleration step. In addition, if $R$ is
larger than zero, it might brake even further in step 2.

The main modification to the NS model takes place in step 3. To determine $%
v_{n}$ consistently for all vehicles, step 3 must be iterated ($v_{max}-1)$
times. We consider the distance between the vehicles $ith$ and $(i+1)th$ and
their coresponding velocities. The knowledge of the velocity of the vehicle
ahead is incorporated through the safety parameter $\alpha $. This parameter
may take values between 0 and 1. If $\alpha $ takes its maximum value, $%
\alpha =1,$ the speed of the vehicle ahead is not considered in the
deceleration process (such as the NS model, which has been compared
succesfully with standard traffic flow in Germany). On the contrary, when $%
\alpha =0$ the speed of the vehicle ahead is considered without
restrictions, i.e., without establishing a safe distance. This fact is
regarding those automated vehicles which can obtain information from the
vehicles in front of them. Thus, the safe distance parameter may allow us to
define the required degree of automation for the system.

This model is a minimal model in the sense that all the four steps are
necessary to reproduce the basic features of real traffic. However,
additional rules may be needed to capture more complex situations.

Thus, the parameters of the model are the following: the limit speed $%
v_{max} $, the braking parameter $R$, the global density $\rho =N/L$ and the
parameter of safe distance $\alpha .$ In our case, $\alpha $ takes the
following values$\ \{0,0.25,0.5,0.75,1\}.$ This latter parameter determines
the safety distances between two vehicles depending of the gap between them
and their speeds. Thus, if $\alpha $ is smaller the restrictions of safety
for the system are lesser and the safety is lower.

\section{Simulation results}

Before going on, we would like to describe our standard simulation set--up
for the following observations. We simulated a system of length $L=10^{4}$
sites with closed boundary conditions, (i.e. traffic was running in a loop)
for a given value of $R$ and $\alpha $.

We start with random initial conditions. $N$ cars are randomly distributed
on the lane around the complete loop with an initial speed taking a discrete
random value between 0 and $v_{max}$. Since the system is closed, the
average density remains constant with time. Next, we update the individual
car velocities and positions in accordance with the rules of the model. Each
density is simulated for $T=6\ast L$ time steps, of which the first half
were discarded to let transients die out and for the system to reach its
asymptotic steady state. Later, we start to extract data to make the
statistical analysis.

For each simulation, we establish a value for parameter $\alpha $ by taking
into account the desired degree of safe distance among vehicles$.$ For
example, the case of $\alpha =0$ is equivalent to taking into account the
distance between two cars and the exact velocity of the vehicle ahead, i.
e., we allow that a vehicle may be behind the other one with the same speed
without the existence of a safe distance between them to prevent accidents.

A convenient way to investigate the model is to draw a diagram of flow
versus density, the so-called fundamental diagram. It is a smooth curve with
a well-defined maximum at a certain density, $\rho _{m}.$ By analyzing the
fundamental diagram, we found that for a given $R,$ the flux exhibits a
maximum at the density $\rho _{m},$\ that decreases when the range of
interaction of the vehicles increases and the length of this interaction is
determined by the safety parameter $\alpha $. In particular, we found that
for $\alpha >0.7,$ the flow and the magnitude of $\rho _{m}$ decrease with
increasing $\alpha $. On the contrary, for all $\alpha \leq 0.7$ the
magnitude of $\rho _{m}$ decreases with decreasing $\alpha $ and the flow
increases. Moreover, smaller $\alpha $ is, i.e. as the safety factor tends
to zero, the longer the length of interaction of the vehicles is and the
flow decreases more rapidly (see Fig. 2).

On the other hand, for $\alpha =0$ and $\alpha >0.7,$ we can distinguish two
different dynamical regions of traffic flow. Flow is linearly increasing
with increase in traffic density (laminar flow, free-flow); beyond $\rho
_{m} $ the flow becomes linearly decreasing in density (back traveling
start-top waves). In the free-flow region all vehicles can move with high
speed close to the speed limit and the average speed converges to the
maximum velocity. However, beyond a certain density, a further increase of
the density leads to a more rapid decay of flow due to the strong length
correlation among vehicles and jams emerge (as in Fig. 2). \ Nevertheless, \
for $0.25\leq \alpha \leq 0.7$ we distinguish an additional mixed flow
region where the flow increases with increasing $\rho $ but the vehicles do
not move with maximum velocity, the speeds of vehicles start decreasing$.$

By analyzing the relationship between velocity and density we observe that
the velocity decreases more rapidly with decreasing $\alpha $ (see Fig. 3).
\ As a consequence the vehicles move rather slowly in this region as
compared to the free-flow region. The flow takes higher values increasing
the highway vehicular capacity due to formation of small platoons of
vehicles with similar velocities to avoid the propagation of jams. \ 

On the other hand, for all $\alpha \leq 1,$ increasing $R$ not only leads to
lesser flow but also lowers $\rho _{m}$ (see Fig 4a). In addition, we found
that for small $R$ and $\alpha >0.7$, the flow has the same behavior, i.e.
the system is insensitive to the braking $R$ (see Fig 4b).

We now examine the higher moments of the velocity distribution. In
particular, we examine the speed variance and its behavior. For all values
of $\alpha $, during both free-flow and congested phases (where the average
speed tends to the maximum speed) the speed variance is negligible. In
particular, for values of $\alpha $ such that there does not exist the
region of mixed flow, i.e.\ $\alpha <0.25$ or $\alpha >0.7$, the speed
variance starts to increse immediately after the free-flow phase. This
variance continues increasing with density until the variance reaches its
maximum value, which occurs after the free-flow phase (as shown in the Fig.
5 for $\alpha =0$). On the other hand, for $0.25\leq \alpha \leq 0.7$ values
we observe a local maximum of the speed variance after the free-flow phase,
but during the mixed flow phase we note a local minimum in the variance
before the maximum flow is reached (see Fig. 6 for $\alpha =0.25$); the
speed variance newly starts to increase and reaches its maximum value after
the density of maximum flow. 

By comparing the maximum flow and the maximum speed variance for $\alpha =0$
and $\alpha =0.25$ (Fig. 5 and 6 respectly) we observe that the maximum flow
for $\alpha =0$ is $12\%$ higher than for $\alpha =0.2$, although the
maximum speed variance at $\alpha =0.25$ is 50\% smaller than $\alpha =0$.
The behavior observed is a consequence of the safety factor we introduced,
i.e, if the safety factor is higher, there exists a smaller propagation of
fluctuactions and prevent variations of speeds by the formation of platoons
or clusters of vehicles.

In general for all values of $\alpha $ the density where the speed increases
until it is maximum value occurs after the maximum flow density, $\rho _{m}$
(the goal of most traffic planners). High speed variance means that
different vehicles in the system have widely varying speeds. It also means
that a vehicle would experience frequent speed changes per trip through out
the system. In turn, this results in higher trip travel time variance. In
reality, this could also increase the probability of traffic accidents. Thus
it seems reasonable to attempt to steer traffic away from the density
immediately after the density of maximum flow.

\section{Summary and conclusion}

In this paper, we have introduced and investigated a statistical model
capable of describing the traffic of automated vehicles accurately,
considering a safety factor for the distance between two vehicles. We
modified an existing CA model to capture interesting characteristics of
traffic flow. Here, we studied the addition of a safe distance parameter, $%
\alpha ,$ to the Nagel-Shreckenberg model to consider the anticipation of a
driver. This modification allows the two vehicles to be as near as possible
by respecting a safe distance, i.e., a driver decides his new velocity by
taking into account the velocity from the driver ahead and the distance that
exists between them.

We think that a suitable safety factor is important to study the behavior of
traffic flow and increasing the vehicular capacity. By the inclusion of the
consideration of the velocity of the vehicle ahead in the process of
deceleration by taking into account a safe distance (i.e., the driver knows
the velocity of the preceding vehicle) we can obtain different correlations
among vehicles that can increase the capacity of vehicles and decrease the
maximum flow. A decrease in the capacity implies an increase in the maximum
flow or the speed variance reaches its maximum value after the region of
mixed flow; whereas for safety factors very big or null (there does not
exist distance of safety) this variance reaches its maximum value
inmediately after the region of free-flow. Thus, the definition of \ a
suitable safety factor may increase the vehicular capacity.

Finally, our approach here was to search for minimal sets of rules which
reproduce certain macroscopic facts. The advantage of this model is that
relations between rules and macroscopic behavior can be more easily
identified; and we also obtain higher computational speed. Thus, the
particle hopping models are a good tool to simulate a big scale traffic flow.

\section{Acknowledgements}

The authors thank to Dr. Vivechana Agerwal and Dr. Mariano L\'{o}pez de Haro
for the critical reading of the paper. This work was partially supported by
DGAPA UNAM under project IN103100.

\textbf{Figure Captions}

Figure 1. A typical configuration in the NS model. The number in the upper
rigth corner is the speed of the vehicle.

Figure 2. Fundamental Diagram for different values of safety factor, and
noise $R=0.4$.

Figure 3. Relationship between mean velocity and density for $R=0.4$ and
different values of $\alpha $

Figure 4a. Fundamental Diagram for safety factor $\alpha =$0.25 and
different values of $R$.

Figure 4b. Fundamental Diagram for different values of the safety parameter $%
\alpha $ and $R=0.2$.

Figure 5. Speed variance versus density for R=0.4, $\alpha =1.$

Figure 6. Speed variance versus density for R=0.4, $\alpha =0.75.$

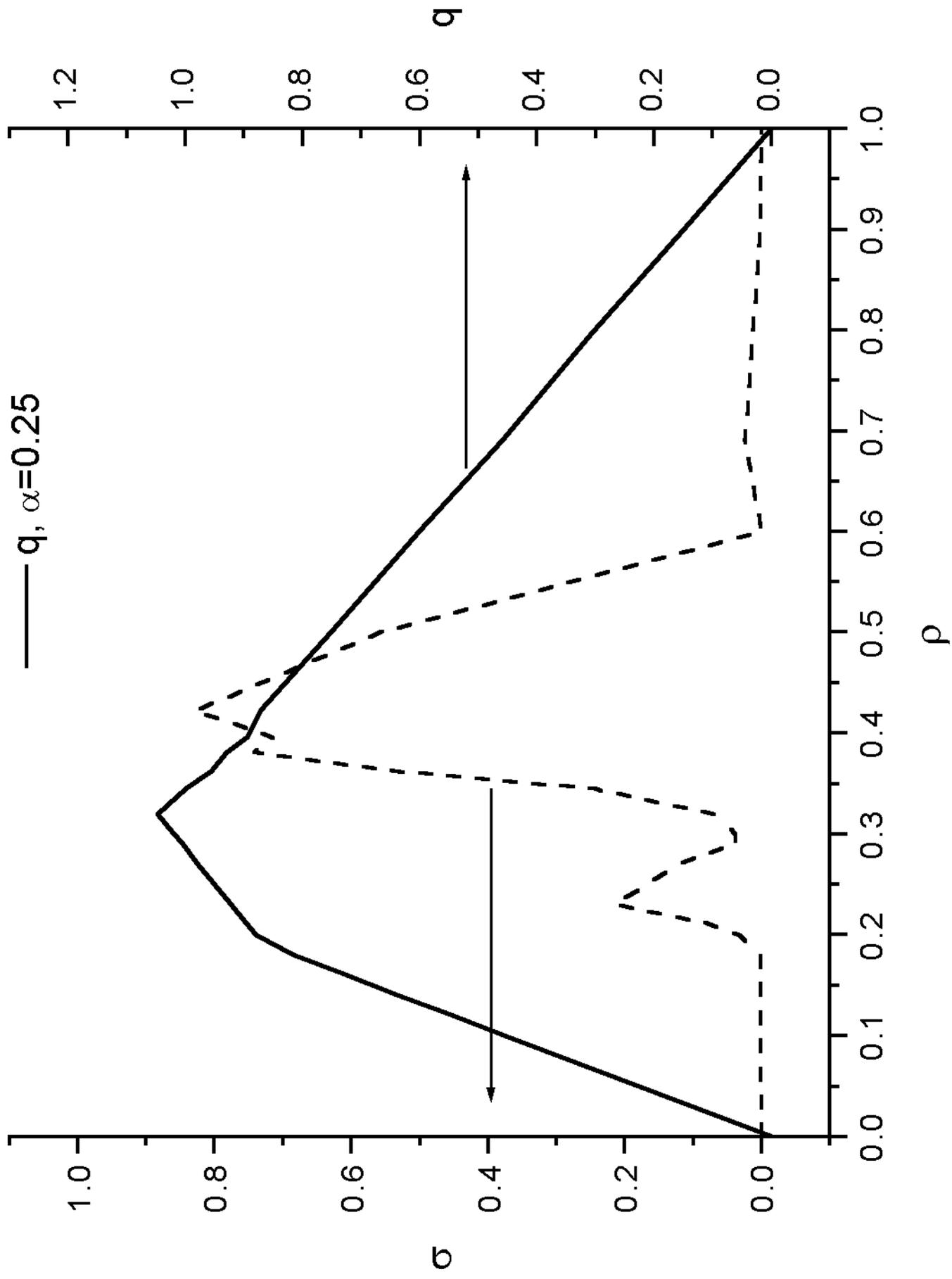
\end{document}